\documentclass[10pt,twocolumn,aps,prl,showpacs,preprintnumbers,amsmath,amssymb,a4paper,superscriptaddress]{revtex4-1}

\usepackage{morefloats,graphicx,color,xkeyval,xparse,xstring}
\usepackage[colorlinks=true,citecolor=black,linkcolor=black,urlcolor=blue,anchorcolor=black]{hyperref}

\usepackage[capitalise]{cleveref}  % must follow after pgfplots, pgfplotstable, or pdf

\definecolor{colspinup}{RGB}{215,48,39}
\definecolor{colspindown}{RGB}{69,117,180}

\DeclareDocumentCommand\opr{m}{\ensuremath{\boldsymbol{\mathbf{#1}}}} % operator syntax
\DeclareDocumentCommand\Tr{r[]}{\ensuremath{\mathrm{Tr}\left[ {#1}\right]}} % trace syntax
\DeclareDocumentCommand\Imag{r[]}{\ensuremath{\mbox{Im}\left[ {#1}\right]}} % imaginary part syntax
\DeclareDocumentCommand\Real{r[]}{\ensuremath{\mbox{Re}\left[ {#1}\right]}} % real part syntax
\DeclareDocumentCommand\log{o}{\ensuremath{\IfNoValueTF{#1}{\mbox{log}}{\mbox{log}\left({#1}\right)}}} % log fn syntax
\def\ct{\dagger}

\DeclareDocumentCommand\degrees{m}{${#1}^{\,\circ}$} % degrees syntax
\def\spup{\uparrow}
\def\spdn{\downarrow}
\def\zz{\ensuremath{\mathrm{zz}}}
\def\ac{\ensuremath{\mathrm{ac}}}
\def\meV{\ensuremath{\text{ meV}}}

\makeatletter
\newcommand{\customlabel}[2]{%
\protected@write\@auxout{}{\string\newlabel{#1}{{\@currentlabel}{\thepage}{#2}{figure.\thepage}{}}}%
\protected@write\@auxout{}{\string\newlabel{#1@cref}{{[figure][\thepage][]\@currentlabel #2}{\thepage}}}%
\hypertarget{#1}{\empty}%
}%
\makeatother

\begin{document}
% ------------------------------------------------------------------------- %

\title{Nanostructured graphene for spintronics}% Force line breaks with \\

\author{S{\o}ren Schou Gregersen}
\email{sorgre@nanotech.dtu.dk}
\affiliation{Center for Nanostructured Graphene (CNG), DTU Nanotech, Department of Micro- and Nanotechnology,
Technical University of Denmark, DK-2800 Kongens Lyngby, Denmark}

\author{Stephen R. Power}
\affiliation{Center for Nanostructured Graphene (CNG), DTU Nanotech, Department of Micro- and Nanotechnology,
Technical University of Denmark, DK-2800 Kongens Lyngby, Denmark}
\affiliation{Department of Physics and Nanotechnology, Aalborg University, Skjernvej 4A, DK-9220 Aalborg East, Denmark}
\affiliation{Catalan Institute of Nanoscience and Nanotechnology (ICN2), CSIC and The Barcelona Institute of Science and Technology, Campus UAB, Bellaterra, E-08193 Barcelona, Spain}
\affiliation{Universitat Autònoma de Barcelona, E-08193 Bellaterra (Cerdanyola del Vallès), Spain}

\author{Antti-Pekka Jauho}
\affiliation{Center for Nanostructured Graphene (CNG), DTU Nanotech, Department of Micro- and Nanotechnology,
Technical University of Denmark, DK-2800 Kongens Lyngby, Denmark}

\date{\today}% It is always \today, today,


% ------------------------------------------------------------------------- %
\begin{abstract}
Zigzag edges of the honeycomb structure of graphene  exhibit magnetic polarization making them  attractive as building blocks for spintronic devices.
Here, we show that devices with zigzag edged triangular antidots perform essential spintronic functionalities, such as spatial spin-splitting or spin filtering of unpolarized incoming currents.
Near-perfect performance can be obtained with optimized structures.
The device performance is robust against substantial disorder.
The gate-voltage dependence of transverse resistance is qualitatively different for spin-polarized and spin-unpolarized devices, and can be used as a diagnostic tool.
Importantly, the suggested devices are feasible within current technologies.
\end{abstract}
% ------------------------------------------------------------------------- %

\maketitle

\textit{Introduction.}
The weak intrinsic spin-orbit coupling and long spin diffusion lengths suggest graphene as an ideal spintronic material~\cite{CastroNeto2009,Han2010,Yazyev2010,Nair2012,McCreary2012,Hong2012,Nair2013,Han2014,Tuan2014,Kamalakar2015}.
% Recent progress has led to efficient spin injection~\cite{Han2010,Friedman2014}, induced  magnetism~\cite{Nair2012,McCreary2012,Hong2012,Nair2013}, and a greater understanding of the mechanisms of spin relaxation~\cite{Han2011,Tuan2014} in graphene.
% A number of theoretical works predict spin splitting or filtering in graphene using, for example, half-metallic nanoribbons~\cite{Son2006,Han2010,Ozaki2010,Saffarzadeh2011}, modulated Rashba fields with strain~\cite{Diniz2017}, flakes~\cite{Sheng2010}, chains~\cite{Zeng2010} or via the Spin Hall Effect (SHE) using graphene with induced topological properties~\cite{Kane2005,Abanin2006,Balakrishnan2014,Cresti2014,Sinova2015}.
Spin splitting or filtering in graphene is predicted for half-metallic nanoribbons~\cite{Son2006,Han2010,Ozaki2010,Saffarzadeh2011}, modulated Rashba fields~\cite{Diniz2017}, flakes~\cite{Sheng2010}, chains~\cite{Zeng2010}, or via the spin Hall effect (SHE)~\cite{Kane2005,Abanin2006,Balakrishnan2014,Cresti2014,Sinova2015}.
Half-metallic systems are excellent platforms for manipulating spin due to their inherent spin filtering behavior.
Self-assembled organometallic frameworks~\cite{Hu2014} and graphene-boron-nitride structures~\cite{Pruneda2010}, point defects and hydrogenation~\cite{Yazyev2007,Palacios2008,Leconte2011}, and, in particular, nanostructured zigzag (zz)-edged devices~\cite{Son2006,Wimmer2008,Wang2009,Zheng2009,Ozaki2010,Sheng2010,Zeng2010,Saffarzadeh2011,Potasz2012,Hong2016,Khan2016,Gregersen2016} are among the proposed graphene-based half metals.
Spin filters have been proposed using triangular dots~\cite{Sheng2010,Hong2016} or perforations~\cite{Zheng2009} with many similarities, e.g., low-energy localized magnetic states and a net sublattice imbalance.
However, perforations, or antidots~\cite{Pedersen2008,Petersen2011,Ouyang2011}, have the advantage over dots of being embedded in the graphene sheet which allows a wide range of spin-dependent transport properties.
Although signatures of localized magnetic states have been detected\cite{Tao2011,Hashimoto2014,Magda2014}, spin manipulation in graphene-based half metals has yet to be realized in experiments.

\begin{figure}[tb]
\centering
\includegraphics{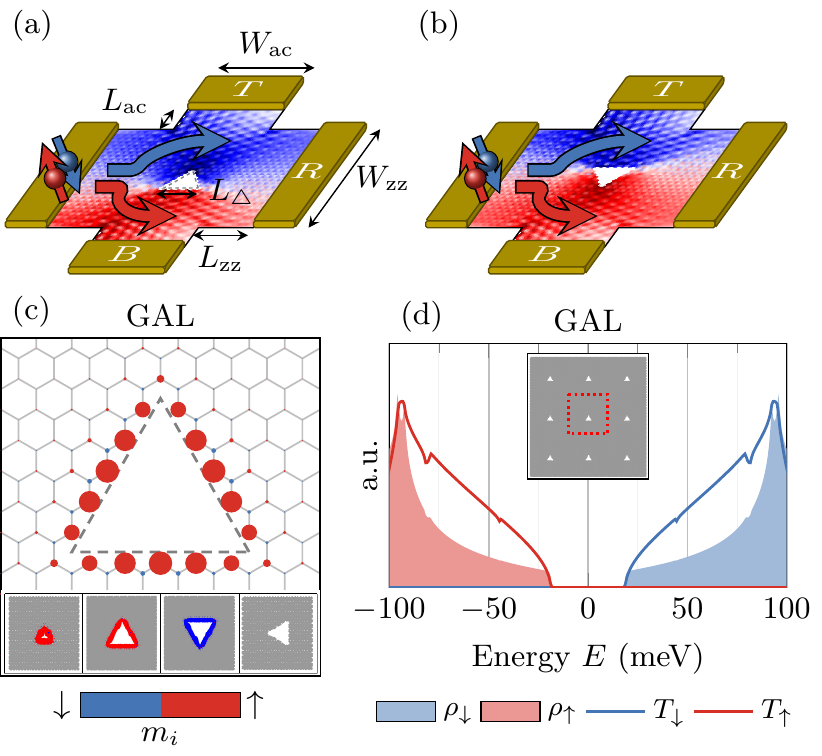}
\caption{
Device geometry (above) and infinite antidot lattice results (below).
Spin up (down) is denoted red (blue) throughout.
(a)
Device geometry, with
$W_\zz \approx 6\text{ nm}$, $W_\ac \approx 4\text{ nm}$, $L_\triangle  \approx 1\text{ nm}$, $L_\zz \approx L_\ac \approx 2\text{ nm}$, and spin-splitting effect with color map displaying the local spin-dependent current magnitude for spin-unpolarized injection through the left electrode (L) $J_{\mathrm{L}}^{\mathrm{s}}$.
(b)
Same as (a) but with a \degrees{180} rotated antidot.
(c)
Magnetic moment profile of the antidot lattice: Moments are represented by circles with radii $\propto |m_i|$.
The inset illustrates magnetic moment profiles for, from left to right, $L_\triangle = 1\text{ nm}$, $L_\triangle = 4\text{ nm}$, rotated $L_\triangle = 4\text{ nm}$, and armchair-edged $L_\triangle = 1\text{ nm}$.
(d)
Spin-dependent transmissions ($T_\sigma$) and density of states ($\rho_\sigma$) of the antidot lattice.
The lattice geometry is shown in the inset and has a $\sim 6 \times 6 \text{ nm}$ square unit cell (dotted box).
}
\customlabel{fig:scheme}{(a)}
\customlabel{fig:scheme.R}{(b)}
\customlabel{fig:moments}{(c)}
\customlabel{fig:dos}{(d)}
\label{fig:scheme_prop}
\end{figure}

In this Rapid Communication, we investigate the transport properties of graphene devices with embedded zz-edged triangular antidots.
Such devices are within the reach of state-of-the-art lithographic methods:
Triangular holes in graphene have recently been fabricated~\cite{Auton2016}, and experiments suggest the possibility of zz-etched nanostructures~\cite{Shi2011,Oberhuber2013}.
Another possibility is to employ a lithographic mask of patterned hexagonal boron nitride, which naturally etches into zz-edged triangular holes~\cite{Jin2009,Gilbert2017}.
The zz-edged structures support local ferromagnetic moments~\cite{Yazyev2010}, however, global ferromagnetism is induced when the overall sublattice symmetry of the edges is broken~\cite{Son2006,Wimmer2008,Wang2009,Guclu2010,Ozaki2010,Zeng2010,Saffarzadeh2011}.
This occurs for zz-edged triangles~\cite{Zheng2009,Sheng2010,Potasz2012,Hong2016,Khan2016,Gregersen2016}.
We have recently discussed the electronic structure of triangular graphene antidot \emph{lattices} (GALs)~\cite{Gregersen2016}—--here, we focus on transport through devices containing a small number of antidots.
Our calculations show that large spin-polarized currents are generated by the device illustrated in \cref{fig:scheme}.
An unpolarized current incident from the left is funneled \emph{below} the triangle if the electron spin is up ($\spup$, red) and \emph{above} if the spin is down ($\spdn$, blue), resulting in spin-polarized currents at contacts top (T) and bottom (B), respectively.

The sixfold symmetry of the graphene lattice allows only two orientations for zz-edged triangles.
A \degrees{180} rotation exposes zz edges with magnetic moments of opposite sign.
In turn, this inverts \emph{both} the scattering directions and spin polarization simultaneously.
An independent inversion of either scattering direction or spin polarization would change the direction of spin current flow, but inverting both restores the spin current flow pattern [\cref{fig:scheme.R}]. 
This results in robust spin behavior over a wide range of superlattice geometries.
The zz-edged triangular GALs have magnetic moment distributions as shown in \cref{fig:moments}, and display half-metallic behavior over a wide range of energies near the Dirac point.
The roles of the two spin orientations can be interchanged by gating, as shown in \cref{fig:dos}.
The magnetic profile remains qualitatively similar when the side length is varied [insets of \cref{fig:moments}], changes sign under a \degrees{180} rotation, and magnetism vanishes for the \degrees{90} rotated (armchair-edged) triangular antidot.
% Kinks within zigzag triangle edges lead to a similar but somewhat reduced magnetic signature~\cite{Yazyev2011}.

In analogy to (inverse) spin Hall measurements~\cite{Sinova2015}, we study the  transverse resistance generated by a longitudinal current.
Using a spin-polarized left contact we suggest a method to distinguish between magnetic or non-magnetic antidots in such devices:
The transverse resistance has a characteristic antisymmetric behavior with respect to the Fermi level only for spin-polarized antidots.

\textit{Geometry and model.}
The device in \cref{fig:scheme} consists of a central graphene region with a single triangular antidot.
(Below we also consider a larger central region with an array of triangles.)
The device has four arms which terminate at metallic contacts---left (L), right (R), top (T), and bottom (B)---which act as sources of either unpolarized or single spin-orientation electrons.
The triangular antidots here have a side length $L_\triangle = 5 a$, where the lattice constant $a = 2.46 \text{ \AA}$.
The remaining dimensions in \cref{fig:scheme} are given in the caption.
Our previous work~\cite{Gregersen2016} validates the use of a nearest-neighbor tight-binding Hamiltonian
$
	\mathcal{H}_{\sigma} = \sum_{i} \epsilon_{i\sigma} {\opr c}_{i\sigma}^{\ct} {\opr c}_{i\sigma} + \sum_{ij} t_{ij} {\opr c}_{i\sigma}^{\ct} {\opr c}_{j\sigma}
$,
to describe the electronic structure of such systems, where ${\opr c}_{i\sigma}^{\ct}$ (${\opr c}_{i\sigma}$) is a creation (annihilation) operator for an electron with spin $\sigma$ on site $i$.
The hopping parameter $t_{ij}$ is $ t = -2.7 \text{ eV}$ for neighbors $i$ and $j$, and zero otherwise.
The T and B arm widths are chosen to yield metallic behavior near the Fermi level $E=0$.

Local magnetic moments are included via spin-dependent on-site energy terms $\epsilon_{i\sigma} = \pm \frac{U}{2} m_i$, with $-$ for $\sigma=\spup$ and $+$ for $\sigma=\spdn$.
The on-site magnetic moments $m_i=\left<{\opr n}_{i\spup}\right> - \left<{\opr n}_{i\spdn}\right>$, where ${\opr n}_{i\sigma}$ is the number operator, are calculated from a self-consistent solution of the Hubbard model within the mean-field approximation.
This is performed for the corresponding extended GAL, displayed in the inset of \cref{fig:dos}, which is an approximately square lattice with a $25 a \times 15 \sqrt{3} a$ ($\sim 6 \text{ nm} \times 6 \text{ nm}$) unit cell.
The four short graphene arm segments are assumed to be nonmagnetic in order to isolate the magnetic influence of the antidots.
An on-site Hubbard parameter $U= 1.33 |t|$ gives results in good agreement with \emph{ab initio} calculations in the case of graphene nanoribbons~\cite{Yazyev2010}.
The sublattice-dependent alignment of moments agrees with Ruderman-Kittel-Kasuya-Yosida (RKKY) theory predictions~\cite{Saremi2007,Power2013}.
Our calculations assume that this extends to inter triangle alignments also.
Due to the large total moment at each triangle, the inter triangle couplings should be stronger than those between, e.g., vacancy defects with similar separations.

The transmission $T_{\alpha\beta}^{\sigma}$ for spin $\sigma$ between two leads $\alpha$ and $\beta$ and local (bond) currents $\opr{J}^\sigma_\alpha$ from lead $\alpha$ are calculated using recursive Green's function techniques~\cite{Lewenkopf2013}.
They are $T_{\alpha\beta}^{\sigma}(E) = \Tr[\opr{\Gamma}_\alpha \opr{G}_\sigma^{r} \opr{\Gamma}_\beta \opr{G}_\sigma^{a}]$ and $\left[\opr{J}^\sigma_\alpha\right]_{ij} = \left[\opr{H}^\sigma\right]_{ji} \Imag[\opr{G}_{\sigma}^{r} \opr{\Gamma}_\alpha \opr{G}_{\sigma}^{a}]_{ij}$, respectively.
$\opr{G}_\sigma^{r}$ ($\opr{G}_\sigma^{a}$) is the retarded (advanced) Green's function, $\opr{\Gamma}_\alpha = -2 \Imag[\opr{\Sigma}_\alpha]$ is the broadening for lead $\alpha$, $\opr{\Sigma}_\alpha$ is the self-energy, and $i$ and $j$ are indices of neighboring sites.
The spin and charge transmissions and local currents are defined for independent spin channels as $T_{\alpha\beta}^{\mathrm{s}}(E) = T_{\alpha\beta}^{\spup}(E) - T_{\alpha\beta}^{\spdn}(E)$, $T_{\alpha\beta}^{\mathrm{c}}(E) = T_{\alpha\beta}^{\spup}(E) + T_{\alpha\beta}^{\spdn}(E)$, $\opr{J}_{\alpha}^{\mathrm{s}} (E) = \opr{J}_{\alpha}^{\spup} (E) - \opr{J}_{\alpha}^{\spdn} (E)$, and $\opr{J}_{\alpha}^{\mathrm{c}} (E) = \opr{J}_{\alpha}^{\spup} (E) + \opr{J}_{\alpha}^{\spdn} (E)$, respectively.
The metallic leads are included via an effective self-energy $\Sigma_{\mathrm{metal}} = - i |t|$ added to the edge sites of the metal/graphene interfaces~\cite{Bahamon2013}.
For spin-polarized contacts, the self-energy for one spin channel is set to zero.
The four-terminal transverse resistance $R_{\mathrm{xy}}$ is determined using L and R as the source and drain and T and B as voltage probes,
\begin{align}
	R_{\mathrm{xy}} = V_{\mathrm{TB}}/I_{\mathrm{L}}^{\mathrm{c}} \,.
\end{align}
where the transverse potential drop $eV_{\mathrm{TB}} = \mu_{\mathrm{T}} - \mu_{\mathrm{B}}$.
Using the Landauer-B\"uttiker relation, the charge currents through lead $\alpha$ are $I_{\alpha}^{\mathrm{c}} = \sum_{\beta\sigma} T_{\beta \alpha}^{\sigma} \left( \mu_{\alpha} - \mu_{\beta} \right)$.
It is assumed that spin mixing occurs in the T and B leads, yielding spin-unpolarized potentials $\mu_{\mathrm{T}}^{\spup} = \mu_{\mathrm{T}}^{\spdn}$ and $\mu_{\mathrm{B}}^{\spup} = \mu_{\mathrm{B}}^{\spdn}$.
We apply source and drain potentials $\mu_{\mathrm{L}} = eV_{\mathrm{LR}}$ and $\mu_{\mathrm{R}} = 0$, while T and B probes carry zero current, $I_{\mathrm{T}}^{\mathrm{c}} = I_{\mathrm{B}}^{\mathrm{c}} = 0$.
The resistance is then determined by solving for $\mu_{\mathrm{T}}$, $\mu_{\mathrm{B}}$, and the longitudinal current.

\begin{figure}
\centering
\includegraphics{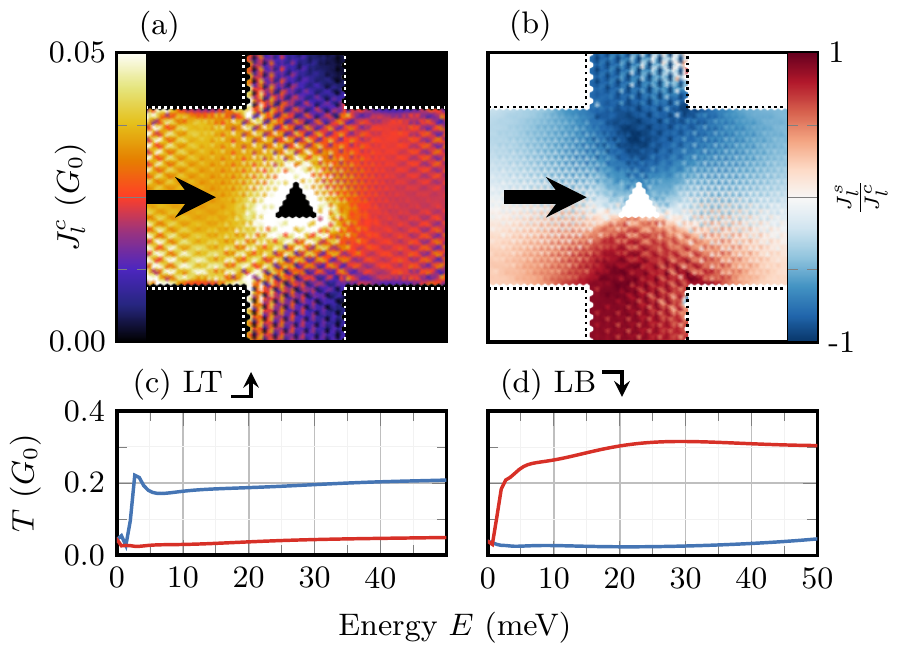}
\caption{
	(a) Local current magnitude through a device with a single triangular antidot at  $E = 20 \text{ meV}$.
	(b) Spin polarization of currents in the same system (red: spin up, blue: spin down).
	Bottom: Spin-dependent transmissions left top [LT, (c)] and left bottom [LB, (d)].
}
\customlabel{fig:1x1transport.currents.charge}{(a)}
\customlabel{fig:1x1transport.currents.spin}{(b)}
\customlabel{fig:1x1transport.LT}{(c)}
\customlabel{fig:1x1transport.LB}{(d)}
\label{fig:1x1transport}
\end{figure}

\textit{Results and discussion.}
Transport properties of the system in \cref{fig:scheme} are presented in \cref{fig:1x1transport}.
The spatial spin separation is illustrated by the magnitude of the local charge current
$J_{\mathrm{L}, i}^{\mathrm{c}} = \left[\opr{J}_{\mathrm{L}}^{\mathrm{c}}\right]_i$
and its spin polarization
$J_{\mathrm{L}, i}^{\mathrm{s}}/J_{\mathrm{L}, i}^{\mathrm{c}}  = \left[\opr{J}_{\mathrm{L}}^{\mathrm{s}}\right]_i/\left[\opr{J}_{\mathrm{L}}^{\mathrm{c}}\right]_i$ at $E=20 \meV$, in \cref{fig:1x1transport.currents.charge,fig:1x1transport.currents.spin} respectively.
At this energy, $\spdn$ electrons are channeled above the antidot and $\spup$ electrons below it.
Incoming $\spup$ electrons are backscattered near the top vertex of the triangular antidot.
This $\spup$-electron behavior is also seen for both spins in the unpolarized system, i.e., letting all $m_i \rightarrow 0$ (not shown), and is due to geometrical factors:
The jagged top half of the device is a more effective backscatterer in general than the nanoribbonlike bottom half.
Conversely, the $\spdn$ behavior is the opposite
Backscattering occurs in the lower half of the device.
This behavior is indicative of scattering near the bottom edge of the triangle which only occurs for $\spdn$ electrons.
This is supported by the presence of strong $\spdn$ local density of states (DOS) features at the middle of each edge in the corresponding bulk lattice~\cite{Gregersen2016}.
Therefore, the scattering of $\spup$ electrons is dictated mainly by the triangular shape of the antidot, whereas $\spdn$ electrons are more sensitive to the magnetic profile. 
The L-T and L-B transmissions shown in \cref{fig:1x1transport.LT,fig:1x1transport.LB} reveal that the spin polarization occurs for a broad range of energies.
Thus, a single-antidot device can partially split or filter incoming currents into either T or B with a large degree of polarizations $T^{\mathrm{s}}/T^{\mathrm{c}}  \sim 70\!\--\!90 \text{ \%}$.

\begin{figure}
\centering
\includegraphics{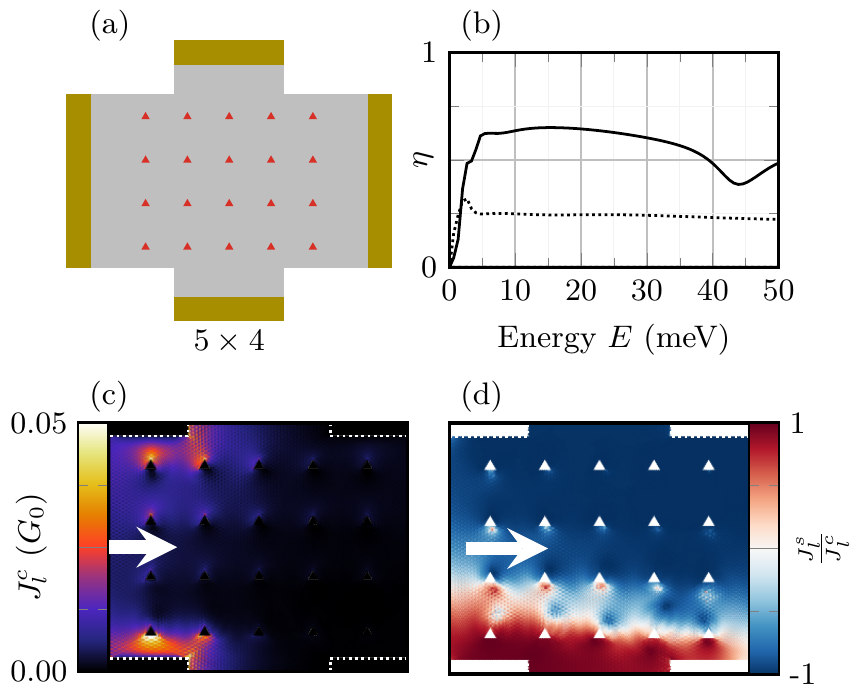}
\caption{
(a)
Schematic of the graphene cross device with a $5 \times 4$ array of triangular antidots.
All triangles are oriented in the same direction with same magnetic moment profile, i.e., as in \cref{fig:moments} (red triangles).
$W_\zz = 60 \sqrt{3} a \approx 26\text{ nm}$ and $W_\ac = 131 a/2 \approx 16\text{ nm}$.
(b)
The splitting efficiency, as defined in main text, of the $5 \times 4$ array  (solid) and the single-antidot  (dotted) devices.
(c) The local charge currents and (d) spin polarization for injection from the left electrode for $E = 20 \text{ meV}$ (red: spin up, blue: spin down).
}
\customlabel{fig:5x4.scheme}{(a)}
\customlabel{fig:5x4transport.efficiency}{(b)}
\customlabel{fig:5x4transport.currents.charge}{(c)}
\customlabel{fig:5x4transport.currents.spin}{(d)}
\label{fig:5x4transport}
\end{figure}
% It is natural to investigate whether larger arrays of antidots can improve the device performance.
A $5 \times 4$ array of triangular antidots is shown in \cref{fig:5x4.scheme}.
We first assume that the magnetic moment profile is the same for each antidot [illustrated in \cref{fig:5x4.scheme} by red triangles] but below we relax this assumption.
The electronic splitting of the spin currents can be quantified by an effective figure of merit
\begin{align}
	\eta &= \frac{T_{\mathrm{LT}}^{\spdn} - T_{\mathrm{LT}}^{\spup} + T_{\mathrm{LB}}^{\spup} - T_{\mathrm{LB}}^{\spdn}}{\sum_{\sigma\alpha\neq\mathrm{L}} T_{\mathrm{L\alpha}}^{\sigma}} \,,
\end{align}
where $\eta \to 1$ for perfect spatial spin splitting into T and B.
The figure of merit in \cref{fig:5x4transport.efficiency} is larger for the array (\emph{solid line}) than for the single-antidot device (\emph{dotted line}), further illustrated by the charge and spin currents at $E = 20 \meV$ in \cref{fig:5x4transport.currents.charge,fig:5x4transport.currents.spin}.
The $\spup$ electrons are effectively blocked away from the array because of half metallicity at this energy, and are either backscattered, or directed towards the B contact.
The $\spdn$ electrons, on the other hand, may enter the array, but have a large probability of deflection towards the T contact due to repeated scattering of the type discussed for the single-antidot case.
Thus, a large imbalance between the spin-resolved transmissions develops, with T and B polarizations  $T^{\mathrm{s}}/T^{\mathrm{c}}  \sim 99 \text{ \%}$ around $E = 20 \text{ meV}$, and $\eta$ is enhanced.

The $\spdn$ behavior is similar to the ratchet effect previously noted for triangular perturbations in graphene~\cite{Koniakhin2014}.
The spatial spin splitting shown here is somewhat analogous to the SHE~\cite{Kane2005,Abanin2006,Balakrishnan2014,Cresti2014,Sinova2015}, where currents of opposite spin are pushed to the opposite edges of the device.
A key distinction is that our device does not require spin-orbit coupling, or topologically protected transport channels.
Even though the antidots share many similarities with regular dots, the enhanced spin splitting by repeated scattering from different antidots is difficult to envision in a dot-based system.

\begin{figure}
\centering
\includegraphics{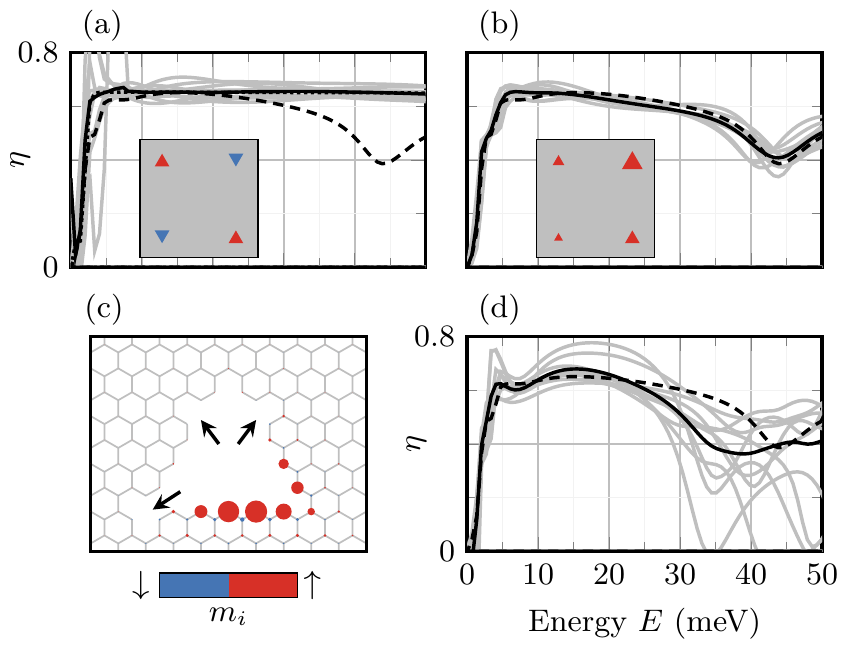}
\caption{
(a, b, d) Transport splitting efficiency of $5 \times 4$ antidot arrays with different disorder.
The splitting efficiency of ten disordered device realizations is shown in gray, the configurational average for each type in black, and the pristine $5 \times 4$ array in black dashed [reproduced from \cref{fig:5x4transport.efficiency}].
(a)
Random flipping of antidots and reversal of spin-polarization (see inset).
An additional array realization with every second antidot flipped ($5 \times 4_{\mathrm{R}}$) is shown by the black dotted curve.
(b)
Random variation of side length, as in the inset.
(c)
Realization of an antidot with removed atoms (black arrows) and the corresponding magnetic moment profile.
The moment profile here and in \cref{fig:moments} are scaled equally.
(d)
Splitting efficiencies for antidot edge atom disorder [see (c)].
}
\customlabel{fig:5x4dtransport.dF}{(a)}
\customlabel{fig:5x4dtransport.dL}{(b)}
\customlabel{fig:5x4dtransport.dE.scheme}{(c)}
\customlabel{fig:5x4dtransport.dE.efficiency}{(d)}
\label{fig:5x4dtransport}
\end{figure}
In experiments, disorder severely degrades properties of atomically precise antidot lattices~\cite{Power2014}.
The half metallicity of triangular GALs is unusually robust against lattice disorder~\cite{Gregersen2016}.
In \cref{fig:5x4dtransport}, we study the effect of disorder in a $5 \times 4$ antidot array using three different methods and ten realizations of each disorder type.

The first disorder type is a random flip of individual antidots, as illustrated in the inset of \cref{fig:5x4dtransport.dF}.
The individual (gray solid) and averaged (black solid) figures of merit for this disorder [\cref{fig:5x4dtransport.dF}] are of the same order as the pristine $5 \times 4$ array (black dashed).
This is expected as the standard and flipped single-triangle devices display very similar behavior [\cref{fig:scheme,fig:scheme.R}].
For comparison, for the case where every second antidot has been flipped [$5 \times 4_{\mathrm{R}}$ (black dotted)], the efficiency is almost exactly identical to the disordered average.
The spread of the different disorder realizations (gray curves) is very small, suggesting that the orientation of the individual antidots plays only a very minor role in these devices, and may even improve the figure of merit compared to the lattice of aligned antidots.

The second disorder type [inset of \cref{fig:5x4dtransport.dL}] randomly varies the triangle side lengths $L \rightarrow L \pm \delta L$, where $\delta L \in \{0, a, 2 a\}$.
The individual and the averaged splitting efficiencies are shown in \cref{fig:5x4dtransport.dL}.
The effect of this disorder is minimal, suggesting that it is the presence of multiple spin-dependent scatterers with similar qualitative behavior and not their exact positioning or size, which enhances the spin-splitting effect.
Enlarging or shrinking a triangle changes the length of the spin-polarized zz edge, and thus the total magnetic moment of an individual triangle [see the inset of \cref{fig:moments}, and the Supplemental Material~\footnote{See Supplemental Material [appended] for magnetic and transport properties of triangular antidots when varying side length and introducing edge disorder.}].
However, the qualitative scattering processes are unchanged.

The third type of disorder, in \cref{fig:5x4dtransport.dE.scheme}, randomly removes $N_{\mathrm{rem}} \leq 3$ edge atoms.
Removing an edge atom splits the zz edges into smaller segments and significantly influences the magnetic moment profile (see also the Supplemental Material~\cite{Note1}).
Random flipping of local moments should play a similar role.
Each device realization comprises of several antidots with a randomly chosen $N_{\mathrm{rem}}\in \{0, 1, 2, 3\}$.
The splitting efficiencies shown in \cref{fig:5x4dtransport.dE.efficiency} show some deviations from pristine behavior.
This can be attributed to the reduction of the total magnetic moment as well as the random introduction of scattering centers at each of the antidots.
Edge disorder is particularly severe for small antidots and is capable of quenching magnetism entirely at some edges.
The longer edge lengths likely in experiment will be more robust against this type of disorder.

\begin{figure}
\centering
\includegraphics{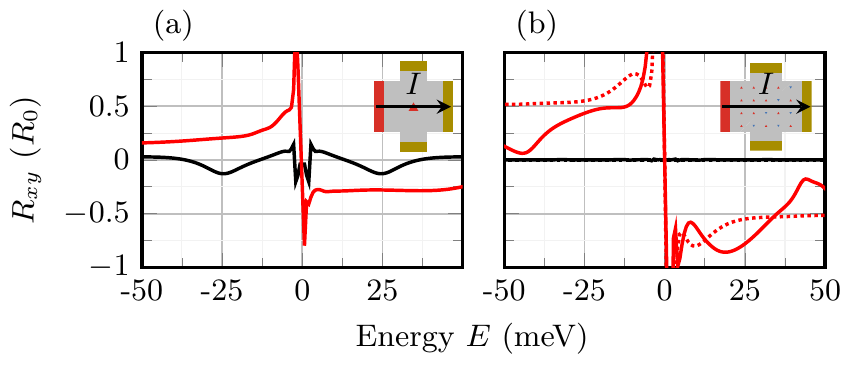}
\caption{
Transverse resistance $R$ in units of $R_0= h / e^2$ for a single antidot device and two $5 \times 4$ array devices when injecting only $\spup$ electrons into the L lead (shown in red).
The resistances of spin-unpolarized ($m_i = 0$) and spin-polarized ($M = \sum m_i \neq 0$) antidot devices are shown in black and red, respectively.
(a)
Single-antidot device, with the inset showing a schematic of $\spup$-polarized electron injection.
(b)
$5 \times 4$ array devices with aligned (solid) and the alternatingly flipped triangles (dotted).
The inset shows the schematic of a $5 \times 4$ array device with $\spup$-polarized electron injection).
}
\customlabel{fig:Rmeasurement.1x1}{(a)}
\customlabel{fig:Rmeasurement.5x4}{(b)}
\label{fig:Rmeasurement}
\end{figure}
Finally, we consider the transverse resistance in a four-terminal device.
The resistances $R_{\mathrm{xy}}$ of the single-antidot device and the $5 \times 4$ and $5 \times 4_{\mathrm{R}}$ devices are shown in \cref{fig:Rmeasurement.1x1,fig:Rmeasurement.5x4}, respectively.
The difference between the top and bottom chemical potentials is $\mu_{\mathrm{T}} - \mu_{\mathrm{B}} \propto T_{\mathrm{L}\mathrm{T}}^{\mathrm{c}} T_{\mathrm{R}\mathrm{B}}^{\mathrm{c}} - T_{\mathrm{R}\mathrm{T}}^{\mathrm{c}}T_{\mathrm{L}\mathrm{B}}^{\mathrm{c}}$, and vanishes in the case of complete left-right symmetry.
For spin-unpolarized electrons the system is exactly L/R symmetric and the resistance is zero (not shown).
\cref{fig:Rmeasurement} shows cases with a $\spup$-polarized L lead.
The transverse resistances in \cref{fig:Rmeasurement.1x1} through a single magnetic antidot (red) show clear antisymmetry with respect to energy.
At positive energies, the fact that the $\spdn$ electrons are now \emph{not} flowing between L and T has the effect of shifting the potential at T closer to that at the R lead, i.e., $\mu_{\mathrm{T}} < eV_{\mathrm{LR}}/2$.
Simultaneously, the potential at B remains close to midway between the L and R potential, i.e., $\mu_{\mathrm{B}} \sim eV_{\mathrm{LR}}/2$.
This yields a negative transverse potential drop $\mu_{\mathrm{T}} - \mu_{\mathrm{B}} < 0$ and in turn a negative resistance $R_{xy} < 0$.
For $E<0$ the spins are flipped and the sign of both the potential drop and the resistance is inverted.
When the antidot is unpolarized positive and negative energies behave similarly, and the resistance is symmetric across the Fermi level, as shown in \cref{fig:Rmeasurement} (black).
% \footnote{there is some nonzero resistance due to spin mixing in the T and B leads}
The same is seen for the both the $5 \times 4$ array and the $5 \times 4_{\mathrm{R}}$ array devices in \cref{fig:Rmeasurement.5x4}.
This clear distinction between magnetic and nonmagnetic antidots provides an excellent measure of whether the device actually splits spin currents, and can, in general, be used to detect magnetism in other nanostructured devices.

\textit{Summary.}
We have demonstrated that magnetic triangular antidots in graphene provide an efficient platform for spatial spin-splitting devices.
The incoming current is split into output leads according to spin orientation, analogous to the spin Hall effect, but without relying on spin-orbit effects.
The outgoing spin polarizations can be flipped using a gate potential.
The predicted performance is robust against typical disorders present in realistic devices.
The transverse resistance yields a clear signal distinguishing the magnetic nature of the perforations.

\begin{acknowledgments}
\textit{Acknowledgments.}
The Center for Nanostructured Graphene (CNG) is sponsored by the Danish National Research Foundation, Project DNRF103.
S.R.P. acknowledges funding from the European Union’s Horizon 2020 research and innovation programme under the Marie Sk\l{}odowska-Curie grant agreement No 665919 and the Severo Ochoa Program (MINECO, Grant No. SEV-2013-0295).
\end{acknowledgments}

\bibliography{references}

%%%%%%%%%% Merge with supplemental materials %%%%%%%%%%
\widetext
\clearpage
\begin{center}
\textbf{\large Nanostructured graphene for spintronics\\Supplemental Material}
\end{center}
%%%%%%%%%% Merge with supplemental materials %%%%%%%%%%
%%%%%%%%%% Prefix a "S" to all equations, figures, tables and reset the counter %%%%%%%%%%
\setcounter{equation}{0}
\setcounter{figure}{0}
\setcounter{table}{0}
\setcounter{page}{1}
\makeatletter
\renewcommand{\theequation}{S\arabic{equation}}
\renewcommand{\thefigure}{S\arabic{figure}}
\renewcommand{\bibnumfmt}[1]{[S#1]}
\renewcommand{\citenumfont}[1]{S#1}
%%%%%%%%%% Prefix a "S" to all equations, figures, tables and reset the counter %%%%%%%%%%

\section{Magnetic moment profiles and disorder}
We consider here the moment profiles of the minimum and maximum side length triangles that can occur occur in the disordered samples in Fig 4 of the main text. 
\begin{figure}[h]
\centering
\includegraphics{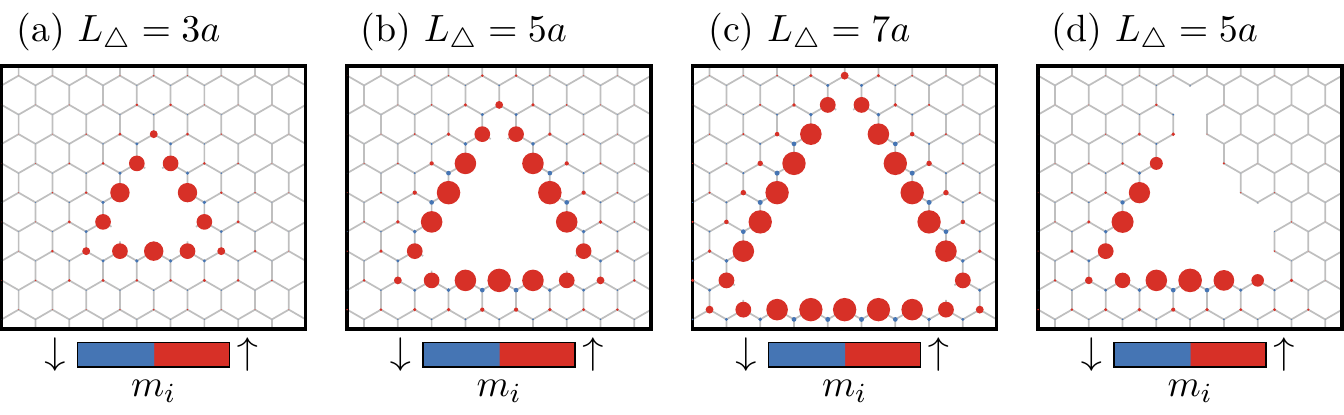}
\caption{
Magnetic moments of the triangular antidot with (a) $L_\triangle = 3a$, (b) $L_\triangle = 5a$, (c) $L_\triangle = 7a$, and (d) $L_\triangle = 5a$ with edge disorder.
The magnetic moments are represented by circles whose radii are scaled by $|m_i|$.
}
\customlabel{fig:supplementmoments.L3}{(a)}
\customlabel{fig:supplementmoments.L5}{(b)}
\customlabel{fig:supplementmoments.L7}{(c)}
\customlabel{fig:supplementmoments.L5.d}{(d)}
\label{fig:supplementmoments}
\end{figure}

As shown in \cref{fig:supplementmoments.L3,fig:supplementmoments.L5,,fig:supplementmoments.L7} these profiles are similar regardless of side length. 
Extended zz-edges yield local profiles resembling those of graphene zz-nanoribbon edges, while corners display reduced profiles.
With edge disorder however, see \cref{fig:supplementmoments.L5.d} where two edge atoms have been removed, the magnetic moment profile may be significantly reduced.
Particularly so if segments become too short to support magnetic states (not shown).
While removing the top corner atom has little influence, the edge atom on the right side reduces the local edge magnetic moments to almost zero.

\section{Transport versus side length}
\begin{figure}
\centering
\includegraphics{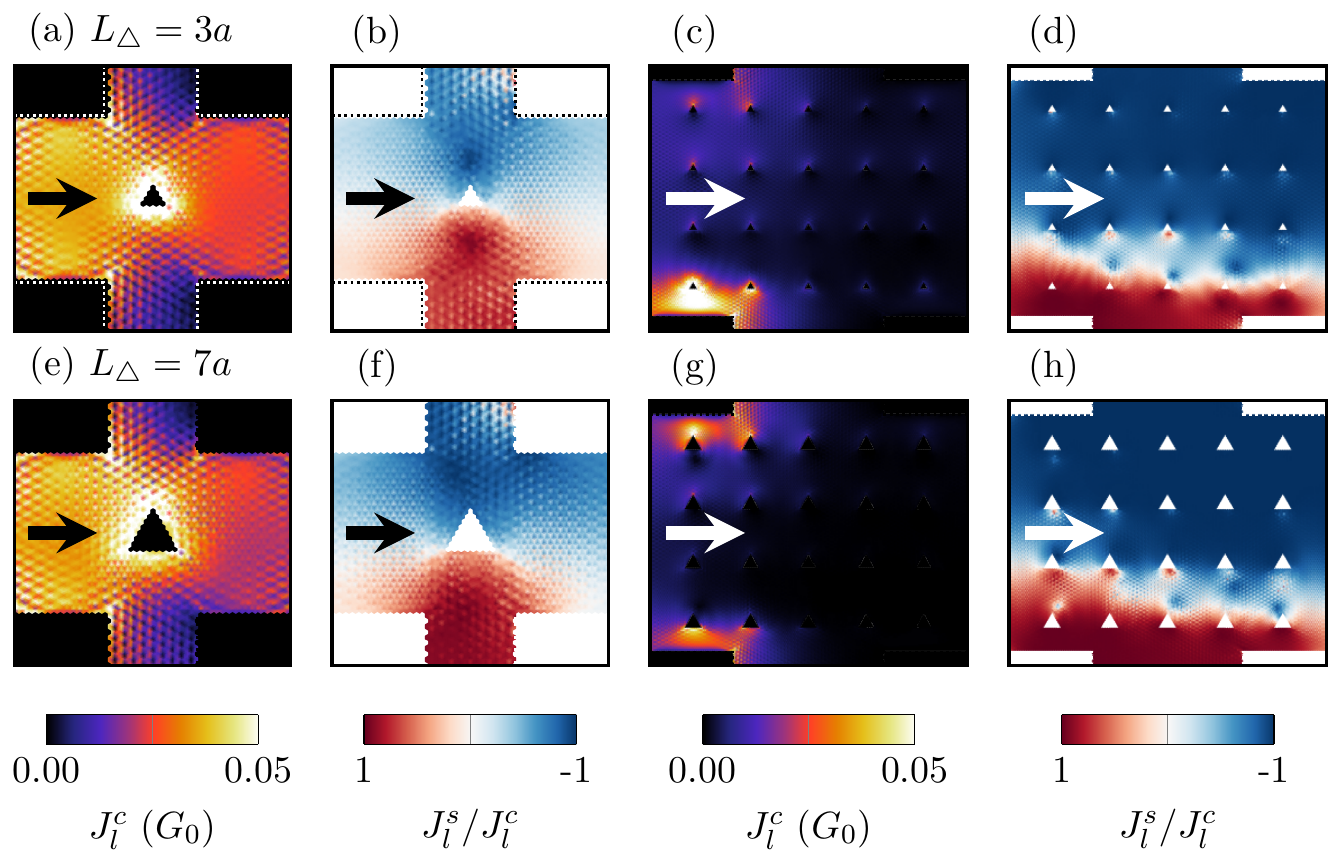}
\caption{
Transport for triangular antidots with (a-d) $L_\triangle = 3a$ and (e-h) $L_\triangle = 7a$.
(a,c,e,g) Local current magnitude through the device at $E = 20 \text{ meV}$.
(b,d,f,h) Spin polarization of currents through the device at $E = 20 \text{ meV}$, with spin up and down transport shown by \emph{red} and \emph{blue}, respectively.
The devices are made from (a,b,e,f) a single antidot or (c,d,g,h) a $5 \times 4$ array of antidots.
}
\customlabel{fig:supplementcurrents.L3.1x1.c}{(a)}
\customlabel{fig:supplementcurrents.L3.1x1.s}{(b)}
\customlabel{fig:supplementcurrents.L3.5x4.c}{(c)}
\customlabel{fig:supplementcurrents.L3.5x4.s}{(d)}
\customlabel{fig:supplementcurrents.L7.1x1.c}{(e)}
\customlabel{fig:supplementcurrents.L7.1x1.s}{(f)}
\customlabel{fig:supplementcurrents.L7.5x4.c}{(g)}
\customlabel{fig:supplementcurrents.L7.5x4.s}{(h)}
\label{fig:supplementcurrents}
\end{figure}

\begin{figure}
\centering
\includegraphics{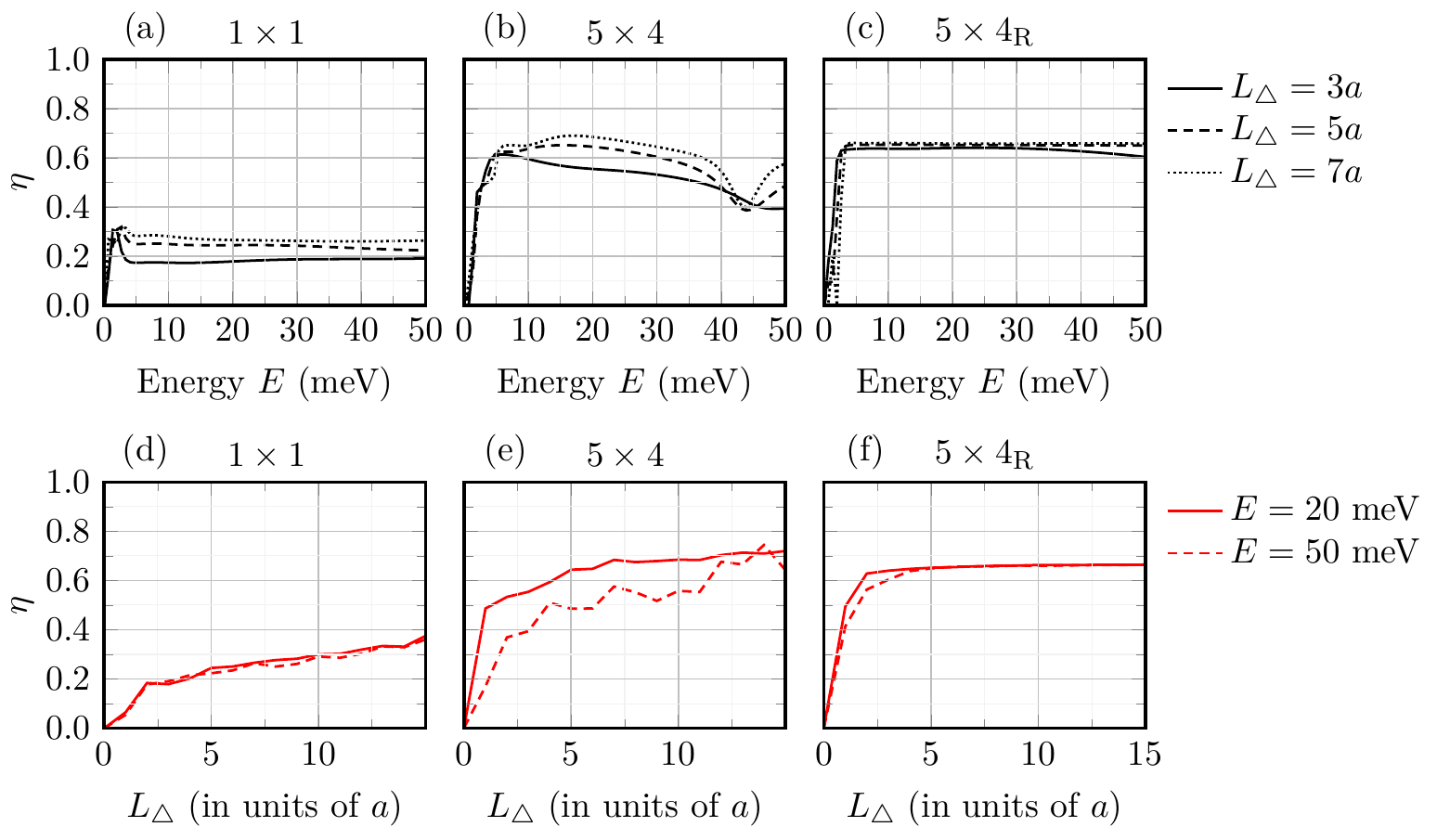}
\caption{
The splitting efficiency of the single and array of antidot devices (a-c) versus energy and (d-f) side length $L_\triangle$.
(a-c) Efficiencies of $L_\triangle = 3a$ (\emph{solid}), $L_\triangle = 5a$ (\emph{dashed}), and $L_\triangle = 7a$ (\emph{dotted}).
(a-c) Efficiencies at $E = 20 \text{ meV}$ (\emph{solid}) and $E = 50 \text{ meV}$ (\emph{dashed}).
}
\customlabel{fig:supplementefficiency.1x1}{(a)}
\customlabel{fig:supplementefficiency.5x4}{(b)}
\customlabel{fig:supplementefficiency.5x4r}{(c)}
\customlabel{fig:supplementefficiency.1x1.varyL}{(d)}
\customlabel{fig:supplementefficiency.5x4.varyL}{(e)}
\customlabel{fig:supplementefficiency.5x4r.varyL}{(f)}
\label{fig:supplementefficiency}
\end{figure}
The electronic transport, governed by the magnetic profiles, also behaves similarly for different side lengths.
In \cref{fig:supplementcurrents}, the $E = 20 \text{ meV}$ currents are displayed using single antidot or arrays antidots of either $L_\triangle = 3 a$ or $L_\triangle = 7 a$.
In comparison the two different sizes yield very similar spin dependent scattering.
Both display the same top-bottom spatial spin-splitting seen in the main text Figs. 2 and 3 with $L_\triangle = 5 a$.

\clearpage
The corresponding spin splitting efficiencies $\eta$ (introduced in the main text) are shown in \cref{fig:supplementefficiency}.
These are illustrated for the single antidot, the $5 \times 4$ array, and the $5 \times 4$ array where every second antidot has been rotated \degrees{180} ($5 \times 4_{\mathrm{R}}$) in \cref{fig:supplementefficiency.1x1,fig:supplementefficiency.5x4,,fig:supplementefficiency.5x4r}, respectively.
Two main results are:
(1) while arrays display enhanced efficiencies, the $5 \times 4_{\mathrm{R}}$ arrays yield the largest efficiencies, and
(2) increasing side lengths also gives larger efficiencies.
Furthermore, in \cref{fig:supplementefficiency.1x1.varyL,fig:supplementefficiency.5x4.varyL,,fig:supplementefficiency.5x4r.varyL} the efficiencies are shown varying side length beyond what is considered in the main text.
At both of the two energies $E = 20 \text{ emV}$ (\emph{solid line}, energy of the current maps) and $E = 50 \text{ emV}$ (\emph{dashed line}), increasing the side length in general increases the efficiencies.
% The limit $\eta \sim 0.67$ seen in \cref{fig:supplementefficiency.5x4r.varyL}, is attributed to a maximum $2/3$ spatial spin splitting and $1/3$ transmission straight through from left to right.

\section{Disordered antidots transport}
\begin{figure}[h]
\centering
\includegraphics{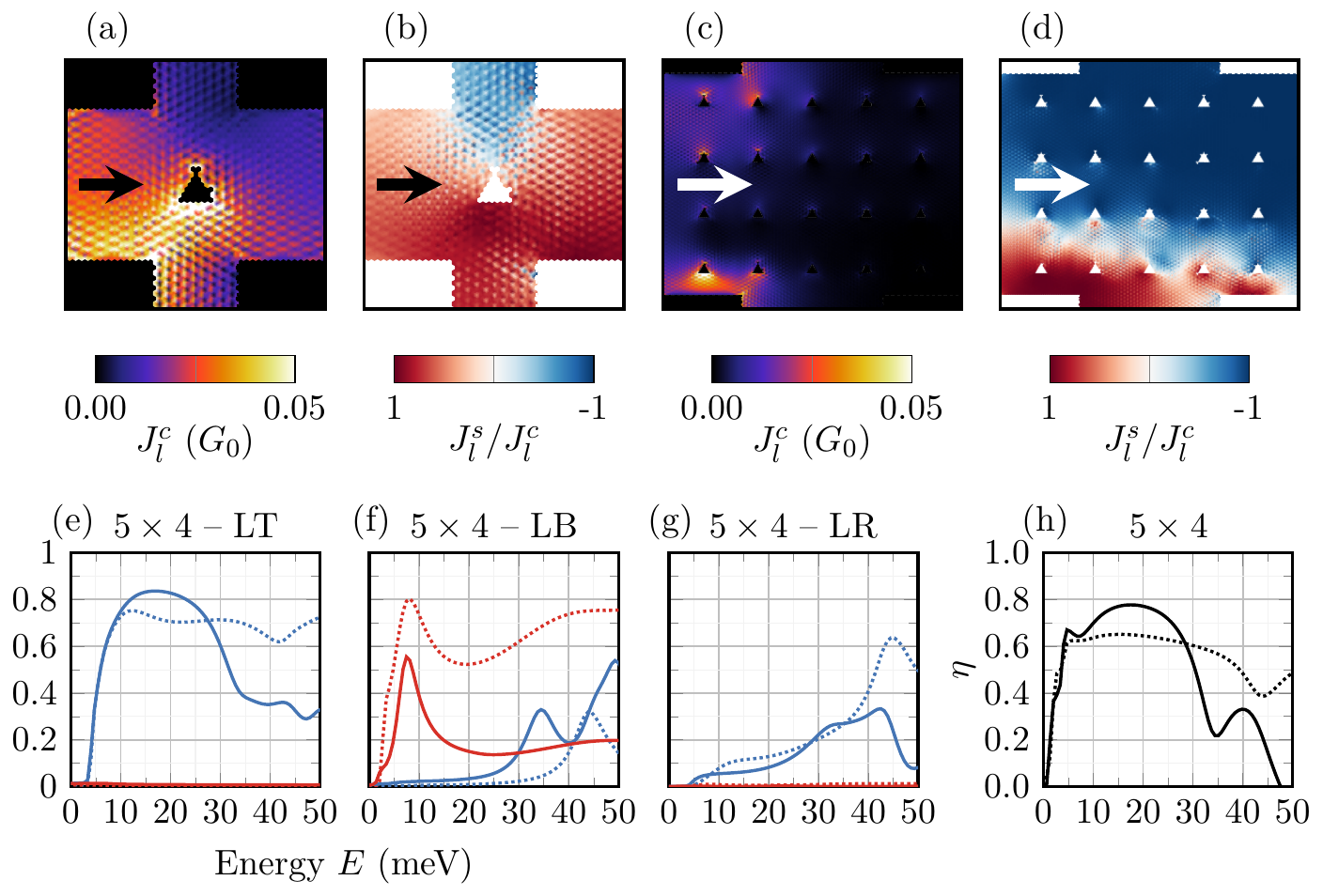}
\caption{
Transport for disordered triangular antidots with $L_\triangle = 5a$.
(a) Local current magnitude through a single antidot device at $E = 20 \text{ meV}$.
(b) Spin polarization of currents through a single antidot device at $E = 20 \text{ meV}$.
(c) Local current magnitude through a $5 \times 4$ array device at $E = 20 \text{ meV}$.
(d) Spin polarization of currents through a $5 \times 4$ array device at $E = 20 \text{ meV}$.
Spin up and down transport shown by \emph{red} and \emph{blue}, respectively.
(e), (f), and (g) show the spin resolved left-top (LT), the left-bottom (LB), and left-right (LR) transmissions, respectively.
(h) The splitting efficiency of the array device in (c) and (d).
}
\customlabel{fig:supplementdisorder.1x1.c}{(a)}
\customlabel{fig:supplementdisorder.1x1.s}{(b)}
\customlabel{fig:supplementdisorder.5x4.c}{(c)}
\customlabel{fig:supplementdisorder.5x4.s}{(d)}
\customlabel{fig:supplementdisorder.5x4.LT}{(e)}
\customlabel{fig:supplementdisorder.5x4.LB}{(f)}
\customlabel{fig:supplementdisorder.5x4.LR}{(g)}
\customlabel{fig:supplementdisorder.5x4.eff}{(h)}
\label{fig:supplementdisorder}
\end{figure}

We now consider transport through devices with edge-disordered triangles.
The transport properties are displayed in \cref{fig:supplementdisorder}.

Even though the splitting of the current is very different in the disordered single antidot case shown in \cref{fig:supplementdisorder.1x1.c,fig:supplementdisorder.1x1.s}, interestingly, the disordered $5 \times 4$ arrays in \cref{fig:supplementdisorder.5x4.c,fig:supplementdisorder.5x4.s} are remarkably similar to the pristine cases of $L_\triangle = 5 a$ in Figs. 2 and 3 of the main text, suggesting conserved spatial spin splitting.
This point is further illustrated with the individual transmissions in \cref{fig:supplementdisorder.5x4.LT,fig:supplementdisorder.5x4.LB,,fig:supplementdisorder.5x4.LR}.
The disordered transmissions (\emph{solid}) also show spin splitting over a wide range of energies, albeit often lower in magnitude compared to the pristine case of $L_\triangle = 5 a$ (\emph{dotted}).
Despite the fluctuations in individual transmissions, in \cref{fig:supplementdisorder.5x4.eff} the splitting efficiencies remain similar to the pristine case across a broad energy range.

\end{document}